\documentclass[sigconf]{acmart}

\usepackage{booktabs}
\usepackage{multirow}
\usepackage{graphicx}
\usepackage{amsmath}
\usepackage{xcolor}
\usepackage{hyperref}
\usepackage{multirow}    
\usepackage{makecell}   
\usepackage{colortbl}
\definecolor{clinred}{RGB}{200,30,30}

\definecolor{clinred}{RGB}{200,30,30}

\AtBeginDocument{%
  }

\setcopyright{rightsretained}
\copyrightyear{2026}
\acmYear{2026}
\acmDOI{}
\acmConference[KDD RelSciFM '26]{Workshop on Reliable Scientific Foundation
  Models}{August 2026}{Jeju, Korea}
\acmISBN{}

\begin{document}

\title{BCoughBench: Benchmarking Respiratory Acoustic Foundation Models
       Under Body-Coupled Wearable Sensor Conditions}


\author{Mayur Sanap}
\affiliation{%
  \institution{Centific Global Solutions Inc.}
  \country{USA}}
\email{mayur.sanap@centific.com}

\author{Prasanna Desikan}
\affiliation{%
  \institution{Centific Global Solutions Inc.}
   \city{}
  \state{}
  \country{USA}}
\email{prasanna.desikan@centific.com} 

\author{Edgar Lobaton}
\affiliation{%
  \institution{North Carolina State University}
  \city{}
  \state{}
  \country{USA}}
\email{ejlobato@ncsu.edu } 

\renewcommand{\shortauthors}{Sanap et al.}

\newcommand{\std}[1]{{\scriptsize$\pm$#1}}

\begin{abstract}
Respiratory acoustic foundation models (FMs) are benchmarked
exclusively on smartphone recordings, yet clinical deployment
increasingly targets body-coupled (BC) wearables whose sensors
attenuate high-frequency content through tissue and bone,
leaving FM reliability uncharacterised. We introduce
BCoughBench, evaluating five FMs (OPERA-CT/CE/GT, HeAR,
M2D+Resp) on nine classification tasks (AUROC, sensitivity at
95\% specificity, Expected Calibration Error) and three age
regression tasks (MAE vs.\ a mean-predictor baseline) across
five EBEN-simulated BC sensor conditions on five labeled cough
datasets. Mean AUROC declines from 0.785 (smartphone) to
0.689--0.723, degrading most under temple vibration pickup
($\Delta$\,=\,$-$0.096) and least under the soft in-ear
($\Delta$\,=\,$-$0.062). No FM meets the clinical sensitivity
threshold (Se@Sp95\,$\geq$\,0.20) on most disease tasks under
any BC sensor. Sex classification on the CIDRZ cohort collapses
(AUROC 0.954\,$\rightarrow$\,0.596--0.628, $\Delta$\,=\,$-$0.341)
while COVID detection is nearly unaffected ($\Delta$\,=\,$-$0.004).
Age regression is robust, improving under the forehead
accelerometer on CoughVID (MAE 9.61\,$\rightarrow$\,8.97\,yr);
HeAR leads on regression and demographic tasks, M2D+Resp on
disease and characteristic tasks. BCoughBench provides a
reproducible framework for FM evaluation under
wearable conditions.
\end{abstract}



\maketitle

\section{Introduction}
\label{sec:intro}

Cough is a cardinal symptom of diseases responsible for
millions of deaths annually, including tuberculosis (TB),
chronic obstructive pulmonary disease (COPD), and COVID-19,
and its acoustic properties encode discriminative signal for
disease detection, demographic inference, and physiological
estimation~\cite{sovijarvi2000respiratory,bhattacharya2023coswara,baur2024hear}.
Foundation models (FMs) pretrained on unlabelled audio
corpora learn task-agnostic embeddings that transfer via
linear probing~\cite{zhang2024opera,baur2024hear,niizumi2025m2d},
reducing the labelled-data burden in clinical audio AI.
The three leading respiratory FM families, OPERA~\cite{zhang2024opera},
HeAR~\cite{baur2024hear}, and M2D+Resp~\cite{niizumi2025m2d},
have been benchmarked on smartphone recordings only and report
AUROC without sensitivity or calibration metrics, concealing
failure at the operating point that deployed screening systems
must use.

Wearable devices increasingly incorporate body-coupled (BC)
microphones for continuous physiological monitoring,
yet body-coupled sensors attenuate high-frequency content
through tissue and bone~\cite{hauret2023eben}, introducing
spectral degradation absent from any current respiratory FM
benchmark. Prior work on body-coupled audio addresses speech
enhancement~\cite{sui2024tramba} and bandwidth
extension~\cite{hauret2024vibravox}, not health inference,
and FM performance under these conditions remains entirely
uncharacterised.

We address this gap with \textbf{BCoughBench}, evaluating
five respiratory FMs across nine classification and three
age regression tasks under five simulated body-coupled
sensor conditions via pre-trained EBEN (Extreme Bandwidth Extension Network) reverse
models~\cite{hauret2023eben,hauret2024vibravox}.
Our contributions are:

\begin{itemize}

\item \textbf{BC simulation pipeline.} EBEN reverse models
applied to five cough datasets produce body-coupled
equivalents across five sensor placements without physical
wearable hardware, with forehead preserving content up to
8\,kHz and throat attenuating above 1.5\,kHz
(Figure~\ref{fig:spectrogram}).

\item \textbf{Multi-metric evaluation.} AUROC, Se@Sp95, and
ECE for classification; MAE vs.\ mean-predictor baseline
for regression. No FM meets Se@Sp95\,$\geq$\,0.20 on most
disease task under any BC sensor.

\item \textbf{Sensor degradation characterisation.} Mean
AUROC drops from 0.785 to 0.689 under temple pickup
($\Delta$\,=\,$-$0.096) and 0.723 under soft in-ear ($\Delta$\,=\,$-$0.062); HeAR and M2D+Resp
retain highest absolute performance.

\item \textbf{Task-dependent findings.} Sex classification
on CIDRZ collapses (AUROC: 0.954\,$\rightarrow$\,0.596--0.628),
COVID detection is nearly unaffected ($\Delta$\,=\,$-$0.004),
and age regression improves under forehead accelerometer
(MAE: 9.61\,$\rightarrow$\,8.97\,yr).

\end{itemize}


\begin{figure*}[t]
  \centering
  \includegraphics[width=\textwidth]{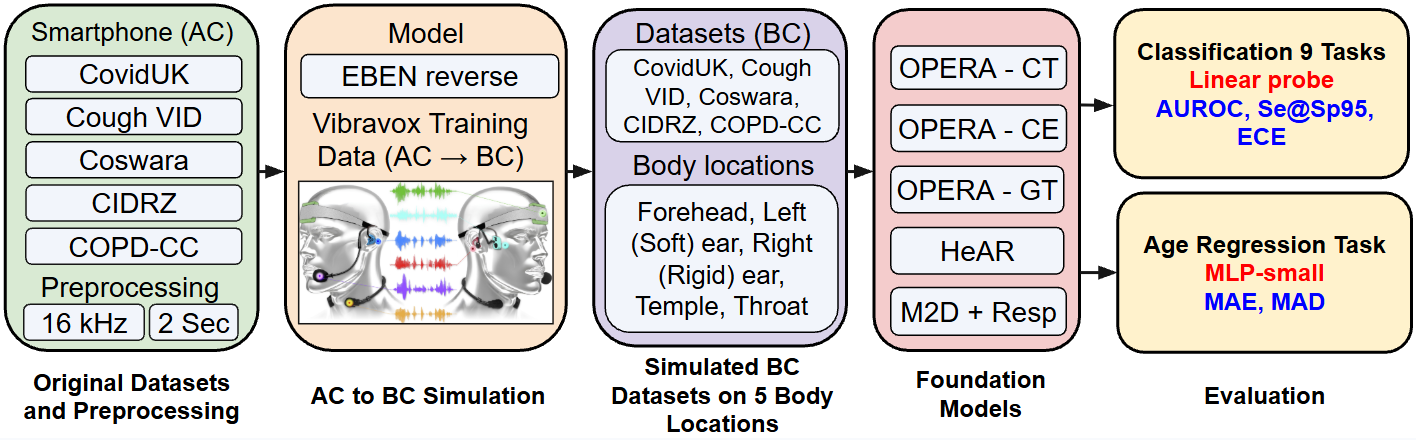}
  \caption{BCoughBench evaluation pipeline. Smartphone cough audio from five
  labeled datasets is preprocessed to 16\,kHz\,/\,2\,s clips and converted to
  body-coupled equivalents via pre-trained EBEN reverse
  models~\cite{hauret2023eben,hauret2024vibravox} across five sensor placements.
  Resulting BC clips are encoded by five frozen respiratory FMs and evaluated
  on nine classification tasks (linear probe; AUROC, Se@Sp95, ECE) and three
  age regression tasks (MLP-small; MAE vs.\ MAD baseline).}
  \label{fig:pipeline}
\end{figure*}

\section{Benchmark Design}
\label{sec:benchmark}

BCoughBench evaluates five respiratory FMs under simulated
body-coupled sensor conditions across nine classification and
three age regression tasks. Figure~\ref{fig:pipeline} illustrates
the full pipeline.

\subsection{Datasets and Tasks}
\label{subsec:datasets}

Table~\ref{tab:datasets} summarises the five source datasets and
twelve BCoughBench tasks spanning disease detection, demographic
classification, acoustic characterisation, and age regression.
All recordings are resampled to 16\,kHz, converted to mono, and
centre-cropped or zero-padded to 2\,s prior to simulation.
Official splits are reused where available, otherwise
subject-disjoint splits are applied.

CoughVID~\cite{orlandic2021coughvid} is a large crowdsourced
cough corpus with self-reported COVID-19 status, acoustic, and
demographic labels collected via smartphone globally.
Coswara~\cite{bhattacharya2023coswara} contains shallow-cough
recordings from India with symptomatic status, sex, cough type,
and age labels.
CIDRZ~\cite{baur2024hear} is a clinical corpus from Zambia with
PCR-confirmed TB labels and demographic annotations.
COPD-CC~\cite{copdcc2025} contains cough recordings from a
Chinese cohort with confirmed COPD diagnosis.
CovidUK~\cite{coppock2021coviduk} contains crowdsourced UK cough
recordings with PCR-confirmed COVID-19 labels.

Task categories are chosen to probe distinct axes of FM embedding
quality under body-coupled degradation. Disease tasks test whether
clinically relevant signal (TB, COPD, symptomatic screening,
COVID-19) survives body-coupled transduction, as these represent
the primary deployment motivation for wearable cough monitoring.
Demographic tasks probe whether sex remains decodable from BC
embeddings, serving as a spurious-correlation indicator.
Characteristic tasks evaluate whether fine-grained acoustic cough
properties (wet/dry, shallow/heavy) are preserved under spectral
degradation; note that the wet/dry task has only 415 samples with
severe class imbalance (356/59) and should be interpreted
cautiously. Regression tasks measure whether physiological signal
(age) survives BC transduction. 

\begin{table}[t]
\caption{Task summary. $N$ = clips per task.
Distribution: class counts (majority\,/\,minority) or
mean\,$\pm$\,std for regression. Audio: 16\,kHz\,/\,2\,s.
Dis.\,=\,Disease, Demo.\,=\,Demographic,
Char.\,=\,Characteristic, Reg.\,=\,Regression.}
\label{tab:datasets}
\small
\setlength{\tabcolsep}{3pt}
\begin{tabular}{llrll}
\toprule
Dataset & Task & $N$ & Cat. & Distribution \\
\midrule
\multirow{3}{*}{CoughVID~\cite{orlandic2021coughvid}}
  & Symptomatic/Healthy & 6,763 & Dis.  & 5628\,/\,1135 \\
  & Wet/Dry cough       &   415 & Char. & 356\,/\,59    \\
  & Age (yr)            & 6,858 & Reg.  & $34.5\pm12.7$ \\
\midrule
\multirow{4}{*}{Coswara~\cite{bhattacharya2023coswara}}
  & Symptomatic/Healthy & 1,983 & Dis.  & 1344\,/\,639  \\
  & Male/Female         & 2,563 & Demo. & 1778\,/\,785  \\
  & Shallow/Heavy cough & 4,992 & Char. & 2496\,/\,2496 \\
  & Age (yr)            & 2,560 & Reg.  & $35.1\pm13.9$ \\
\midrule
\multirow{3}{*}{CIDRZ~\cite{baur2024hear}}
  & TB/non-TB           & 1,049 & Dis.  & 876\,/\,173   \\
  & Male/Female         & 1,049 & Demo. & 535\,/\,514   \\
  & Age (yr)            & 1,049 & Reg.  & $37.1\pm12.9$ \\
\midrule
COPD-CC~\cite{copdcc2025}
  & COPD/Healthy        &   853 & Dis.  & 221\,/\,632   \\
\midrule
CovidUK~\cite{coppock2021coviduk}
  & COVID/Non-COVID     & 2,500 & Dis.  & 1616\,/\,884  \\
\bottomrule
\end{tabular}
\end{table}

\begin{figure*}[t]
  \centering
  \includegraphics[width=\textwidth]{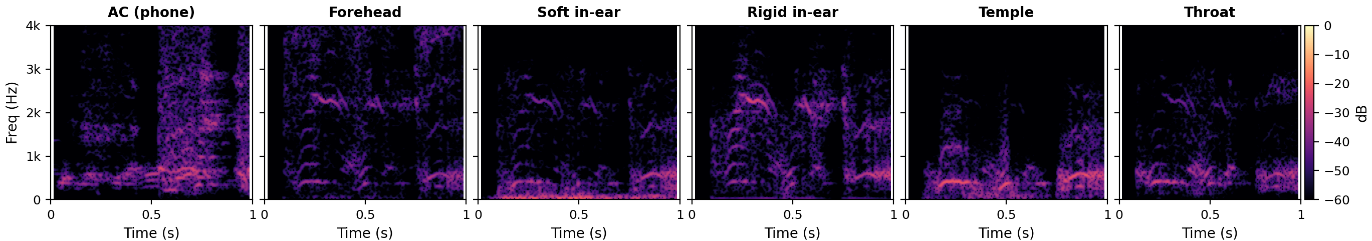}
  \caption{Spectrograms of a representative CoughVID clip
  under AC (smartphone) and five simulated BC sensor conditions
  (1\,s window of peak cough activity, 0--4\,kHz shown).
  High-frequency content is progressively attenuated from
  forehead accelerometer to throat microphone, confirming
  the expected low-pass characteristics of body-coupled
  transduction~\cite{hauret2023eben}.}
  \label{fig:spectrogram}
\end{figure*}

\subsection{AC-to-BC Simulation}
\label{subsec:simulation}

We simulate body-coupled degradation using pre-trained EBEN
reverse models~\cite{hauret2023eben, hauret2024vibravox}.
EBEN~\cite{hauret2023eben}
is a GAN-based architecture using a configurable multiband
decomposition with a U-Net-like generator to recover
high-frequency content lost through body-coupled transduction.
Body-coupled sensors act as low-pass filters with
sensor-dependent cutoff frequencies; the in-ear microphone,
for example, exhibits near-zero coherence with the
air-conducted (AC) signal above 3\,kHz~\cite{hauret2023eben}.
Hauret et al.\ note that this family of sensors degrades
speech in a consistent manner independent of content, with
variations occurring mainly in cutoff frequency and
attenuation~\cite{hauret2023eben}, 
though this content-independence assumption is unverified for impulsive cough.
The Vibravox corpus~\cite{hauret2024vibravox}
provides 45 hours of paired air-conducted and body-coupled
French speech from 188 speakers recorded simultaneously
across six sensor placements, on which EBEN is trained in
both the BC-to-AC enhancement and AC-to-BC simulation
directions. We use the published AC-to-BC checkpoints,
which map clean smartphone audio to each sensor's
characteristic spectral profile, without further adaptation.

We use the term body-coupled as an umbrella covering bone
conduction, tissue conduction, accelerometry, and laryngeal
sensing, since not all five sensors are strictly
bone-conduction devices~\cite{hauret2024vibravox}. The five
sensor placements and their approximate usable bandwidths
are: forehead accelerometer (skull vibration, smart glasses,
$\leq$\,8\,kHz), soft in-ear microphone (occluded ear canal,
left ear, earbuds, $\leq$\,6\,kHz), rigid in-ear microphone
(occluded ear canal, right ear, earbuds, $\leq$\,5\,kHz),
temple vibration pickup (bone conduction, glasses frame,
$\leq$\,2\,kHz), and throat microphone (laryngeal
conduction, throat collar, $\leq$\,1.5\,kHz).

For each sensor $s \in \mathcal{S}$, the corresponding
reverse model transforms each AC clip as
$\hat{x}^{(s)} = G_s(x_{\text{AC}})$, where the output
is zero-padded to 2\,s (32,000 samples at 16\,kHz) and
peak-normalised. 
No BC-specific normalisation or augmentation is applied
at any stage. Figure~\ref{fig:spectrogram} confirms the
expected progressive high-frequency attenuation; validation against real BC cough remains future work.

\subsection{Foundation Models}
\label{subsec:models}

We evaluate five frozen respiratory FMs spanning three
pretraining paradigms. All models receive 2\,s clips at
16\,kHz and are kept frozen throughout; embeddings are
extracted once and reused across all tasks and sensor
conditions.

OPERA-CT and OPERA-CE~\cite{zhang2024opera} are contrastive
models trained on 136K respiratory clips, differing in
architecture (Transformer vs.\ EfficientNet-B0 CNN) and
embedding dimension (768-d vs.\ 1280-d).
OPERA-GT~\cite{zhang2024opera} is a generative masked
autoencoder on the same corpus using an 8.18\,s positional
grid with zero-padded inputs, producing 384-d embeddings.
HeAR~\cite{baur2024hear} is a ViT-L masked autoencoder
pretrained on 313M health audio clips (512-d), the largest
training corpus among the five models and the only one
extending beyond respiratory sounds to broader clinical audio.
M2D+Resp~\cite{niizumi2025m2d} combines masked spectrogram
modelling on AudioSet with respiratory fine-tuning,
producing 3840-d embeddings per clip.

\subsection{Evaluation Protocol}
\label{subsec:protocol}

All five FMs are evaluated in a strictly zero-shot regime;
none has been trained or fine-tuned on body-coupled audio,
and no BC-specific adaptation is applied at any stage.

\textbf{Classification.} Following the linear evaluation
protocol of prior respiratory FM
benchmarks~\cite{zhang2024opera, baur2024hear}, we train a
single linear probe on top of frozen FM embeddings for each
task--sensor combination using the Adam optimiser
($\text{lr}=10^{-4}$, $\ell_2=10^{-5}$, 64 epochs,
batch size 64) with exponential learning rate decay
($\gamma=0.97$). Results are reported as mean $\pm$ std
over 5 random seeds. We report three complementary metrics:
AUROC (discrimination), Se@Sp95 (clinical sensitivity at
95\% specificity), and ECE (calibration). Se@Sp95\,$<$\,0.20
is treated as clinically unusable.

\textbf{Regression.} Age regression uses an MLP-small head
(one hidden layer, 256-unit bottleneck, 0.3 dropout) with
early stopping (patience\,=\,10) monitored on validation
MAE, reported alongside the mean-predictor baseline (MAD)
over 5 seeds. MAE\,$>$\,MAD indicates the model performs
worse than predicting the training mean.

\textbf{Smartphone baselines.} For each task we report the
best-FM smartphone AUROC or MAE as the reference point.
Delta values ($\Delta$\,=\,BC\,$-$\,phone) are reported
for all results; negative $\Delta$ indicates degradation
for classification and improvement for regression.

\begin{table*}[t]
\caption{BCoughBench classification results (mean, 5 seeds).
\textbf{Bold} = best AUROC per row.
\colorbox{red!15}{Red} = Se@Sp95 (SE)\,$<$\,0.20 (clinically unusable).
$^\dagger$ = any model ECE\,$>$\,0.10 at that sensor.
FM: CT\,=\,OPERA-CT, CE\,=\,OPERA-CE, GT\,=\,OPERA-GT,
HR\,=\,HeAR, M2\,=\,M2D+Resp.
$^a$Coswara, $^b$CoughVID, $^c$CIDRZ.
Mean $\Delta$ = mean AUROC across five BC sensors $-$ phone.}
\label{tab:classification}
\small
\renewcommand{\arraystretch}{1.15}
\setlength{\tabcolsep}{4pt}
\begin{tabular}{l cc cc cc cc cc cc c}
\toprule
& \multicolumn{2}{c}{Phone}
  & \multicolumn{2}{c}{Forehead}
  & \multicolumn{2}{c}{Soft-ear}
  & \multicolumn{2}{c}{Rigid-ear}
  & \multicolumn{2}{c}{Temple}
  & \multicolumn{2}{c}{Throat}
  & Mean \\
\cmidrule(lr){2-3}\cmidrule(lr){4-5}\cmidrule(lr){6-7}
\cmidrule(lr){8-9}\cmidrule(lr){10-11}\cmidrule(lr){12-13}
Task
  & AUC & SE
  & AUC & SE
  & AUC & SE
  & AUC & SE
  & AUC & SE
  & AUC & SE
  & $\Delta$ \\
\midrule
\multicolumn{14}{l}{\textit{Disease}} \\
TB/non-TB
  & 0.648 & 0.251
  & 0.591$^\dagger$ (HR) & \cellcolor{red!15}0.074
  & 0.559 (HR) & \cellcolor{red!15}0.057
  & \textbf{0.618} (M2) & \cellcolor{red!15}0.091
  & 0.578 (CE) & \cellcolor{red!15}0.069
  & 0.578 (M2) & \cellcolor{red!15}0.080
  & $-$0.063 \\
COPD/Healthy
  & 0.832 & 0.324
  & 0.805 (M2) & 0.387
  & \textbf{0.825} (M2) & 0.462
  & 0.822 (M2) & 0.427
  & 0.733$^\dagger$ (HR) & 0.258
  & 0.814 (M2) & 0.440
  & $-$0.032 \\
Symptomatic/Healthy$^a$
  & 0.846 & 0.517
  & 0.769 (M2) & 0.388
  & 0.780 (M2) & 0.388
  & \textbf{0.786} (M2) & 0.408
  & 0.749 (M2) & 0.310
  & 0.787 (M2) & 0.346
  & $-$0.073 \\
Symptomatic/Healthy$^b$
  & 0.647 & \cellcolor{red!15}0.124
  & \textbf{0.615} (HR) & \cellcolor{red!15}0.108
  & 0.613 (HR) & \cellcolor{red!15}0.108
  & 0.610 (HR) & \cellcolor{red!15}0.114
  & 0.582 (HR) & \cellcolor{red!15}0.087
  & 0.603 (M2) & \cellcolor{red!15}0.100
  & $-$0.043 \\
COVID/Non-COVID
  & 0.697 & \cellcolor{red!15}0.191
  & \textbf{0.703} (M2) & \cellcolor{red!15}0.180
  & 0.697 (M2) & \cellcolor{red!15}0.187
  & 0.685 (M2) & \cellcolor{red!15}0.163
  & 0.684 (M2) & \cellcolor{red!15}0.179
  & 0.698 (M2) & \cellcolor{red!15}0.169
  & $-$0.004 \\
\midrule
\multicolumn{14}{l}{\textit{Demographic}} \\
Male/Female$^a$
  & 0.924 & 0.741
  & \textbf{0.934} (HR) & 0.764
  & 0.908 (HR) & 0.703
  & 0.887 (M2) & 0.566
  & 0.871 (HR) & 0.586
  & 0.899 (M2) & 0.597
  & $-$0.024 \\
Male/Female$^c$
  & 0.954 & 0.872
  & 0.609 (M2) & \cellcolor{red!15}0.094
  & 0.619 (CE) & \cellcolor{red!15}0.094
  & 0.611 (GT) & \cellcolor{red!15}0.094
  & 0.596 (GT) & \cellcolor{red!15}0.101
  & \textbf{0.628} (CE) & \cellcolor{red!15}0.103
  & $-$0.341 \\
\midrule
\multicolumn{14}{l}{\textit{Characteristic}} \\
Wet/Dry cough
  & 0.711 & \cellcolor{red!15}0.118
  & 0.687 (CT) & \cellcolor{red!15}0.082
  & \textbf{0.732}$^\dagger$ (HR) & \cellcolor{red!15}0.165
  & 0.666 (HR) & \cellcolor{red!15}0.106
  & 0.638 (HR) & \cellcolor{red!15}0.082
  & 0.635$^\dagger$ (CT) & \cellcolor{red!15}0.059
  & $-$0.039 \\
Shallow/Heavy cough
  & 0.809 & 0.339
  & \textbf{0.782} (M2) & 0.297
  & 0.777 (M2) & 0.283
  & 0.771 (HR) & 0.269
  & 0.771 (M2) & 0.303
  & 0.777 (M2) & 0.278
  & $-$0.034 \\
\midrule
\rowcolor{gray!10}
\textbf{Mean (9)}
  & 0.785 & 0.386
  & \textbf{0.722} & 0.264
  & 0.723 & 0.272
  & 0.717 & 0.249
  & 0.689 & 0.219
  & 0.713 & 0.241
  & $-$0.073 \\
\bottomrule
\end{tabular}
\end{table*}

\section{Results}
\label{sec:results}

Full classification results with mean\,$\pm$\,std are in
Tables~\ref{tab:app_auroc}--\ref{tab:app_ece}
(Appendix~\ref{app:full_results}) and full regression
results in Table~\ref{tab:app_reg}
(Appendix~\ref{app:reg_results}).


\subsection{Classification Tasks}
\label{sec:cls_results}

Table~\ref{tab:classification} reports AUROC, Se@Sp95, and ECE
across nine classification tasks and five BC sensors relative
to the smartphone baseline.

\textbf{Overall degradation is moderate but consistent.}
Mean AUROC drops from 0.785 to 0.689--0.723 across
BC sensors. Sensor severity from worst to best:
temple\,$<$\,throat\,$<$\,rigid\,$<$\,soft\,$\approx$\,forehead.

\textbf{Disease detection remains clinically unusable.}
Se@Sp95 $<$ 0.20 on TB, Symptomatic (CoughVID), and
COVID across all five sensors. COPD and Symptomatic
(Coswara) retain clinical sensitivity
(Se@Sp95\,=\,0.258--0.462 and 0.310--0.408).
Best-FM calibration is acceptable (ECE\,$<$\,0.10) except
COPD under temple (0.117); the $\dagger$ markers in
Table~\ref{tab:classification} flag sensors where any model
exceeds 0.10 (TB forehead, Wet/Dry soft in-ear).

\textbf{Sex classification collapses on CIDRZ.}
Male/Female on CIDRZ drops from 0.954 to 0.596--0.628
($\Delta$\,=\,$-$0.341), the largest degradation of any
task, suggesting sex-discriminative features lie in the
high-frequency range destroyed by BC transduction.
By contrast, Male/Female on Coswara improves under
forehead (0.924\,$\rightarrow$\,0.934,
$\Delta$\,=\,$+$0.010), indicating dataset-specific
spectral encoding of sex.

\textbf{COVID detection is uniquely robust.}
COVID/Non-COVID shows near-zero degradation
($\Delta$\,=\,$-$0.004), yet Se@Sp95 remains below 0.20
across all sensors and is clinically unusable.


\textbf{HeAR and M2D+Resp dominate.} M2D+Resp leads on disease and characteristic tasks; HeAR on demographic tasks. OPERA models
underperform both, with OPERA-GT reaching
Se@Sp95\,=\,0.000 on TB on smartphone audio.

\textbf{Wet/Dry improves under soft in-ear.}
Wet/Dry AUROC rises to 0.732 ($+$0.02 vs.\ phone),
suggesting the occluded ear canal amplifies resonance
differences masked in open-air recordings. 



\begin{table}[t]
\caption{Age regression MAE (yr, best FM per sensor, mean
over 5 seeds). \textbf{Bold} = best BC sensor per dataset.
MAD = mean-predictor baseline. All best FM: HeAR.}
\label{tab:regression}
\small
\renewcommand{\arraystretch}{1.15}
\setlength{\tabcolsep}{3pt}
\begin{tabular}{lcccccccc}
\toprule
Dataset & MAD & Phone
  & Forehead & Soft-ear & Rigid-ear & Temple & Throat \\
\midrule
CoughVID & 10.13 & 9.61
  & \textbf{8.97}
  & 9.14
  & 9.20
  & 9.55
  & 9.15 \\
Coswara  & 10.94 & 9.12
  & \textbf{9.07}
  & 9.39
  & 9.41
  & 9.71
  & 9.27 \\
CIDRZ    & 10.42 & 10.29
  & \textbf{10.27}
  & 10.27
  & 10.28
  & 10.29
  & 10.27 \\
\bottomrule
\end{tabular}
\end{table}

\subsection{Regression Tasks}
\label{sec:reg_results}

Table~\ref{tab:regression} reports age MAE across five BC
sensors alongside the smartphone and MAD baselines.
HeAR is the best FM across all sensors and datasets,
consistent with its large and diverse pretraining data.

\textbf{Age regression is broadly robust.} All results
remain well below MAD across all sensors (mean MAE
8.97--10.29\,yr vs.\ MAD 10.13--10.94\,yr), confirming
physiological age signal is preserved under BC
transduction, in stark contrast to classification.

\textbf{CoughVID improves under BC sensors.} All five
sensors yield lower MAE than smartphone (9.61\,yr),
with forehead achieving 8.97\,yr ($-$0.64\,yr), suggesting
low-frequency age correlates are partially masked by
high-frequency noise in open-air recordings.

\textbf{Coswara shows modest degradation on severe sensors.}
Forehead closely matches smartphone (9.07 vs.\ 9.12\,yr)
while temple degrades more noticeably (9.71\,yr),
consistent with its narrow usable bandwidth
($\leq$\,2\,kHz) among the five sensors.

\textbf{CIDRZ is effectively unchanged.} All sensors
yield MAE within 0.02\,yr of smartphone (10.29\,yr),
indicating that age signal in this clinical TB cohort
is contained in low frequencies preserved by all BC
sensors. The robustness of CIDRZ age regression across
all five sensors suggests that clinical cough cohorts
may be more BC-compatible for regression than
crowdsourced datasets.

\section{Discussion}
\label{sec:discussion}
BCoughBench reveals a consistent gap between smartphone FM
performance and body-coupled wearable conditions. No FM
meets Se@Sp95\,$\geq$\,0.20 on most disease tasks under any
BC sensor, a failure invisible to AUROC alone, arguing for
multi-metric reporting as a minimum standard for respiratory
FM evaluation. Sensor selection proves as important as model
selection: the gap between the best sensor (soft in-ear,
mean AUROC\,=\,0.723) and worst (temple, 0.689) is 0.034
points, comparable in magnitude to inter-FM differences on
smartphone audio. Task category determines degradation
severity: sex classification on CIDRZ collapses
($\Delta$\,=\,$-$0.341) while age regression improves under
forehead ($-$0.64\,yr on CoughVID), suggesting that
low-frequency physiological signals survive BC transduction
while high-frequency discriminative cues do not. 

\section{Conclusion}
\label{sec:conclusion}

 
We introduced BCoughBench, evaluating five respiratory FMs
across nine classification and three age regression tasks
under five simulated body-coupled sensor conditions. Mean
AUROC drops from 0.785 to 0.689--0.723 across BC sensors,
no FM meets the clinical sensitivity threshold on most
disease task, sex classification on CIDRZ collapses
($\Delta$\,=\,$-$0.341), and age regression improves under
the forehead accelerometer ($-$0.64\,yr on CoughVID).
HeAR and M2D+Resp retain the highest absolute performance;
sensor selection is as consequential as model selection
for wearable deployment. These results highlight that
AUROC alone is insufficient; Se@Sp95 and ECE are necessary
to surface deployment-critical failures. BCoughBench uses
simulated rather than real BC audio, which may not fully
capture sensor-specific noise or physiological artifacts.
Future work should validate on physical wearable hardware,
extend to additional health targets such as BMI and disease
severity, and explore BC-aware FM pretraining and
adaptation strategies.


\bibliographystyle{ACM-Reference-Format}
\bibliography{bcoughbench_refs}

\newpage
\appendix

\section{Full Classification Results}
\label{app:full_results}

Tables~\ref{tab:app_auroc}, \ref{tab:app_se}, and
\ref{tab:app_ece} report mean\,$\pm$\,std over 5 seeds
for AUROC, Se@Sp95, and ECE respectively for all nine
classification tasks across five BC sensors (best FM
per sensor).

\begin{table*}[h!]
\caption{AUROC mean\,$\pm$\,std (best FM per sensor, 5 seeds).
Phone = single-seed mean (std unavailable).
\textbf{Bold} = best sensor per task.}
\label{tab:app_auroc}
\normalsize
\renewcommand{\arraystretch}{1.2}
\setlength{\tabcolsep}{3pt}
\begin{tabular}{lcccccc}
\toprule
Task & Phone & Forehead & Soft-ear & Rigid-ear & Temple & Throat \\
\midrule
\multicolumn{7}{l}{\textit{Disease}} \\
TB/non-TB         & 0.648 & 0.591\std{0.035} & 0.559\std{0.023} & \textbf{0.618}\std{0.022} & 0.578\std{0.004} & 0.578\std{0.038} \\
COPD/Healthy      & 0.832 & 0.805\std{0.006} & \textbf{0.825}\std{0.006} & 0.822\std{0.007} & 0.733\std{0.005} & 0.814\std{0.008} \\
Symptomatic (Cos) & 0.846 & 0.769\std{0.003} & 0.780\std{0.004} & \textbf{0.786}\std{0.003} & 0.749\std{0.007} & 0.787\std{0.003} \\
Symptomatic (Cov) & 0.647 & \textbf{0.615}\std{0.004} & 0.613\std{0.004} & 0.610\std{0.009} & 0.582\std{0.003} & 0.603\std{0.006} \\
COVID             & 0.697 & \textbf{0.703}\std{0.002} & 0.697\std{0.004} & 0.685\std{0.004} & 0.684\std{0.003} & 0.698\std{0.004} \\
\midrule
\multicolumn{7}{l}{\textit{Demographic}} \\
Male/Female (Cos)  & 0.924 & \textbf{0.934}\std{0.003} & 0.908\std{0.006} & 0.887\std{0.003} & 0.871\std{0.006} & 0.899\std{0.001} \\
Male/Female (CIDRZ)& 0.954 & 0.609\std{0.025} & 0.619\std{0.002} & 0.611\std{0.008} & 0.596\std{0.013} & \textbf{0.628}\std{0.037} \\
\midrule
\multicolumn{7}{l}{\textit{Characteristic}} \\
Wet/Dry cough     & 0.711 & 0.687\std{0.058} & \textbf{0.732}\std{0.038} & 0.666\std{0.033} & 0.638\std{0.027} & 0.635\std{0.016} \\
Shallow/Heavy     & 0.809 & \textbf{0.782}\std{0.001} & 0.777\std{0.002} & 0.771\std{0.002} & 0.771\std{0.001} & 0.777\std{0.002} \\
\bottomrule
\end{tabular}
\end{table*}

\begin{table*}[h!]
\caption{Se@Sp95 mean\,$\pm$\,std (best FM per sensor, 5 seeds).
\colorbox{red!15}{Red} = Se@Sp95\,$<$\,0.20 (clinically unusable).}
\label{tab:app_se}
\normalsize
\renewcommand{\arraystretch}{1.2}
\setlength{\tabcolsep}{3pt}
\begin{tabular}{lcccccc}
\toprule
Task & Phone & Forehead & Soft-ear & Rigid-ear & Temple & Throat \\
\midrule
\multicolumn{7}{l}{\textit{Disease}} \\
TB/non-TB         & 0.251 & \cellcolor{red!15}0.074\std{0.039} & \cellcolor{red!15}0.057\std{0.040} & \cellcolor{red!15}0.091\std{0.033} & \cellcolor{red!15}0.069\std{0.014} & \cellcolor{red!15}0.080\std{0.033} \\
COPD/Healthy      & 0.324 & 0.387\std{0.023} & 0.462\std{0.017} & 0.427\std{0.026} & 0.258\std{0.033} & 0.440\std{0.017} \\
Symptomatic (Cos) & 0.517 & 0.388\std{0.008} & 0.388\std{0.015} & 0.408\std{0.020} & 0.310\std{0.042} & 0.346\std{0.015} \\
Symptomatic (Cov) & \cellcolor{red!15}0.124 & \cellcolor{red!15}0.108\std{0.009} & \cellcolor{red!15}0.108\std{0.007} & \cellcolor{red!15}0.114\std{0.010} & \cellcolor{red!15}0.087\std{0.005} & \cellcolor{red!15}0.100\std{0.008} \\
COVID             & \cellcolor{red!15}0.191 & \cellcolor{red!15}0.180\std{0.012} & \cellcolor{red!15}0.187\std{0.007} & \cellcolor{red!15}0.163\std{0.018} & \cellcolor{red!15}0.179\std{0.005} & \cellcolor{red!15}0.169\std{0.007} \\
\midrule
\multicolumn{7}{l}{\textit{Demographic}} \\
Male/Female (Cos)  & 0.741 & 0.764\std{0.006} & 0.703\std{0.016} & 0.566\std{0.025} & 0.586\std{0.030} & 0.597\std{0.023} \\
Male/Female (CIDRZ)& 0.872 & \cellcolor{red!15}0.094\std{0.042} & \cellcolor{red!15}0.094\std{0.016} & \cellcolor{red!15}0.094\std{0.013} & \cellcolor{red!15}0.101\std{0.013} & \cellcolor{red!15}0.103\std{0.011} \\
\midrule
\multicolumn{7}{l}{\textit{Characteristic}} \\
Wet/Dry cough     & \cellcolor{red!15}0.118 & \cellcolor{red!15}0.082\std{0.047} & \cellcolor{red!15}0.165\std{0.044} & \cellcolor{red!15}0.106\std{0.044} & \cellcolor{red!15}0.082\std{0.029} & \cellcolor{red!15}0.059\std{0.053} \\
Shallow/Heavy     & 0.339 & 0.297\std{0.004} & 0.283\std{0.011} & 0.269\std{0.015} & 0.303\std{0.005} & 0.278\std{0.016} \\
\bottomrule
\end{tabular}
\end{table*}

\begin{table*}[h!]
\caption{ECE mean\,$\pm$\,std (best FM per sensor, 5 seeds).
$^\dagger$ = any model ECE\,$>$\,0.10 at this sensor.}
\label{tab:app_ece}
\normalsize
\renewcommand{\arraystretch}{1.2}
\setlength{\tabcolsep}{3pt}
\begin{tabular}{lcccccc}
\toprule
Task & Phone & Forehead & Soft-ear & Rigid-ear & Temple & Throat \\
\midrule
\multicolumn{7}{l}{\textit{Disease}} \\
TB/non-TB         & 0.055 & 0.0290\std{0.0199} & 0.0459\std{0.0233} & 0.0362\std{0.0169} & 0.0453\std{0.0453} & 0.0584\std{0.0230} \\
COPD/Healthy      & 0.052 & 0.0529\std{0.0085} & 0.0579\std{0.0176} & 0.0564\std{0.0193} & 0.1165$^\dagger$\std{0.0085} & 0.0714\std{0.0197} \\
Symptomatic (Cos) & 0.050 & 0.0652\std{0.0065} & 0.0562\std{0.0172} & 0.0523\std{0.0043} & 0.0533\std{0.0100} & 0.0444\std{0.0107} \\
Symptomatic (Cov) & 0.021 & 0.0353\std{0.0122} & 0.0170\std{0.0035} & 0.0361\std{0.0134} & 0.0258\std{0.0048} & 0.0255\std{0.0049} \\
COVID             & 0.071 & 0.0602\std{0.0049} & 0.0599\std{0.0037} & 0.0585\std{0.0054} & 0.0627\std{0.0082} & 0.0597\std{0.0049} \\
\midrule
\multicolumn{7}{l}{\textit{Demographic}} \\
Male/Female (Cos)  & 0.035 & 0.0437\std{0.0046} & 0.0482\std{0.0150} & 0.0354\std{0.0077} & 0.0679\std{0.0068} & 0.0367\std{0.0050} \\
Male/Female (CIDRZ)& 0.194 & 0.0584\std{0.0135} & 0.0415\std{0.0089} & 0.0540\std{0.0163} & 0.0354\std{0.0116} & 0.0446\std{0.0189} \\
\midrule
\multicolumn{7}{l}{\textit{Characteristic}} \\
Wet/Dry cough     & 0.048 & 0.0820\std{0.0990} & 0.0541$^\dagger$\std{0.0151} & 0.0579\std{0.0251} & 0.0367\std{0.0061} & 0.0278$^\dagger$\std{0.0138} \\
Shallow/Heavy     & 0.028 & 0.0412\std{0.0035} & 0.0357\std{0.0064} & 0.0386\std{0.0052} & 0.0341\std{0.0057} & 0.0485\std{0.0119} \\
\bottomrule
\end{tabular}
\end{table*}

\section{Full Regression Results}
\label{app:reg_results}

Table~\ref{tab:app_reg} reports mean\,$\pm$\,std over 5
seeds for age regression MAE across five BC sensors.
HeAR is the best FM across all sensors and datasets.

\begin{table*}[h!]
\caption{Age regression MAE\,$\pm$\,std (yr, HeAR, 5 seeds).
\textbf{Bold} = best BC sensor per dataset.
MAD = mean-predictor baseline.}
\label{tab:app_reg}
\normalsize
\renewcommand{\arraystretch}{1.2}
\setlength{\tabcolsep}{4pt}
\begin{tabular}{lccccccc}
\toprule
Dataset & MAD & Phone & Forehead & Soft-ear & Rigid-ear & Temple & Throat \\
\midrule
CoughVID & 10.13 & 9.61
  & \textbf{8.97}\std{0.04}
  & 9.14\std{0.01}
  & 9.20\std{0.03}
  & 9.55\std{0.04}
  & 9.15\std{0.04} \\
Coswara  & 10.94 & 9.12
  & \textbf{9.07}\std{0.03}
  & 9.39\std{0.04}
  & 9.41\std{0.03}
  & 9.71\std{0.04}
  & 9.27\std{0.04} \\
CIDRZ    & 10.42 & 10.29
  & \textbf{10.27}\std{0.03}
  & 10.27\std{0.02}
  & 10.28\std{0.03}
  & 10.29\std{0.03}
  & 10.27\std{0.05} \\
\bottomrule
\end{tabular}
\end{table*}

\end{document}